# Spectral decomposition approach to macroscopic parameters of Fokker-Planck flows: Part 2

*Igor A. Tanski*
*Moscow, Russia*
*tanski.igor.arxiv@gmail.com*

*ABSTRACT*

In this paper we proceed with investigation of connections between Fokker - Planck equation and continuum mechanics. We base upon expressions from our work [2], based upon the spectral decomposition of Fokker - Planck equation solution. In this decomposition we preserve only terms with the smallest degrees of damping. We find, that macroscopic parameters of Fokker-Planck flows, obtained in this way, satisfy the set of conservation laws of classic hydrodynamics. The expression for stresses (30) contains additional term - this term is negligible in big times limit. We proved also, that the velocities field alone satisfy Burgers equation without mass forces - but with some additional term. This term is also negligible in big times limit. For the zero degree theory, considered in [1], there are no additional terms. But this theory is valid only for the potential velocities field, fully deductible from density - the potential is proportional to density logarithm. In this theory we can not specify initial conditions for velocities independently from density. Taking in account of the next degree terms could partly solve this problem, but result in some loss of exactness.

This paper is the second part of our work [1]. The aim of these papers is to investigate connections between Fokker - Planck equation and continuum mechanics. We follow the method, proposed in [4]. Namely, in spectral decomposition of solution we preserve only terms with the smallest degree of damping. In the first part we considered only zero degree terms. Here we discuss results of the next degree terms consideration. In the course of calculations we use results of our work [3].

We know from [1], that for zero degree terms satisfy the set of classic hydrodynamics equations for isothermal compressible fluid with friction mass force, proportional to velocity. Velocities alone satisfy in this case the Burgers equation without mass forces. But all this is true only for the potential velocities field, fully deductible from density - the potential is proportional to density logarithm. This is, of course, very special class of flows. In particular, we can not specify initial conditions for velocities independently from density.

We expect, that the taking in account of he next degree terms could solve this problem.

All references to formulae [1] we shall write according to following format: (P1-##), where ## is the placeholder for corresponding formula number.

## 1. Density

To get the first approximation of Fokker - Planck equation solution we keep terms with $p_i = 1$ in (P1-1) - (P1-3).

$$\rho_1 = \int\limits_{-\infty}^{+\infty}\int\limits_{-\infty}^{+\infty}\int\limits_{-\infty}^{+\infty} \exp\left[-t\left(\alpha + \frac{k}{\alpha^2}\sum_{j=1}^{3}\omega_j^2\right)\right] \times \quad (1)$$

$$\times A_{\omega_1\omega_2\omega_3 100}\left(\frac{k}{2\pi\alpha}\right)^{3/2}\left(\frac{2k}{\alpha}\right)^{1/2}\left(\frac{i\omega_1}{\alpha}\right)\prod_{j=1}^{j=3}\exp(i\omega_j x_j)\exp\left[-\frac{k}{2\alpha}\left(\frac{\omega_j}{\alpha}\right)^2\right]d\omega_1 d\omega_2 d\omega_3.$$

It is useful to express $\rho_i$ as a derivative of some potential $\phi_i$ to simplify handling terms with negative degrees of $\omega_j$ in following expressions (see, for example (7)). Besides that we write $e^{-\alpha t}$ factor explicitly to have clear insight of damping velocities of different terms in following expressions. We write

$$\rho_1 = e^{-\alpha t}\frac{\partial \phi_1}{\partial x_1}; \quad (2)$$

where

$$\phi_1 = \int\limits_{-\infty}^{+\infty}\int\limits_{-\infty}^{+\infty}\int\limits_{-\infty}^{+\infty} \exp\left[-t\frac{k}{\alpha^2}\sum_{j=1}^{3}\omega_j^2\right] \times \quad (3)$$

$$\times A_{\omega_1\omega_2\omega_3 100}\left(\frac{k}{2\pi\alpha}\right)^{3/2}\left(\frac{2k}{\alpha}\right)^{1/2}\left(\frac{1}{\alpha}\right)\prod_{j=1}^{j=3}\exp(i\omega_j x_j)\exp\left[-\frac{k}{2\alpha}\left(\frac{\omega_j}{\alpha}\right)^2\right]d\omega_1 d\omega_2 d\omega_3.$$

We see, that potential $\phi_1$ satisfy diffusion equation

$$\frac{\partial \phi_1}{\partial t} = \frac{k}{\alpha^2}\left(\frac{\partial^2 \phi_1}{\partial x^2} + \frac{\partial^2 \phi_1}{\partial y^2} + \frac{\partial^2 \phi_1}{\partial z^2}\right), \quad (4)$$

and density satisfy diffusion with damping equation

$$\frac{\partial \rho_1}{\partial t} + \alpha\rho_1 = \frac{k}{\alpha^2}\left(\frac{\partial^2 \rho_1}{\partial x^2} + \frac{\partial^2 \rho_1}{\partial y^2} + \frac{\partial^2 \rho_1}{\partial z^2}\right). \quad (5)$$

Two another functions $\rho_2, \rho_3$ are defined in the similar to (2) way by potentials $\phi_2$ and $\phi_3$. These potentials satisfy the same equation (4), densities satisfy (5). Let us now denote the previously considered density (P1-8) as $\rho_0$. Then full density $\rho$ for first degree approximation can be expressed as a sum

$$\rho = \rho_0 + \rho_1 + \rho_2 + \rho_3 = \rho_0 + e^{-\alpha t} \frac{\partial \phi_k}{\partial x_k}. \tag{6}$$

Here and in the following repeated indices are understood to be summed (Einstein's summation convention).

## 2. Velocities

We have following expressions for velocities

$$u_{1k} = \frac{i}{\rho} \int_{-\infty}^{+\infty} \int_{-\infty}^{+\infty} \int_{-\infty}^{+\infty} \exp\left[-t\left(\alpha + \frac{k}{\alpha^2}\sum_{j=1}^{3}\omega_j^2\right)\right] \times \tag{7}$$

$$\times A_{\omega_1\omega_2\omega_3 100} \left(\frac{k}{2\pi\alpha}\right)^{3/2} \left[-\frac{k}{\alpha}\frac{\omega_k}{\alpha} - \frac{\alpha\delta_{1k}}{\omega_1}\right] \left(\frac{2k}{\alpha}\right)^{1/2} \left(\frac{i\omega_1}{\alpha}\right) \prod_{j=1}^{j=3} \exp(i\omega_j x_j) \exp\left[-\frac{k}{2\alpha}\left(\frac{\omega_j}{\alpha}\right)^2\right] d\omega_1 d\omega_2 d\omega_3.$$

where $\rho$ is defined by (6). Using this definition, we have

$$u_{1k} = e^{-\alpha t} \frac{1}{\rho}\left(-\frac{k}{\alpha^2}\frac{\partial^2 \phi_1}{\partial x_1 \partial x_k} + \alpha\delta_{1k}\phi_1\right). \tag{8}$$

Recall expression (P1-11) for velocities

$$u_{0k} = \frac{1}{\rho}\left(-\frac{k}{\alpha^2}\frac{\partial \rho_0}{\partial x_k}\right) \tag{9}$$

and have

$$u_k = u_{0k} + u_{1k} + u_{2k} + u_{3k}, \tag{10}$$

or

$$u_k = \frac{1}{\rho}\left(-\frac{k}{\alpha^2}\frac{\partial \rho}{\partial x_k} + \alpha e^{-\alpha t}\phi_k\right). \tag{11}$$

## 3. The continuity equation

As we mentioned in [1], though we dropped all fast damping terms, (6) and (11) are still derived from exact solution of Fokker - Planck equation. Therefore they must satisfy certain conservation laws.

Let us check the continuity equation.

$$\frac{\partial(\rho u_k)}{\partial x_k} = \left(\frac{\partial(\rho u_{0k})}{\partial x_k} + \frac{\partial(\rho u_{1k})}{\partial x_k} + \frac{\partial(\rho u_{2k})}{\partial x_k} + \frac{\partial(\rho u_{3k})}{\partial x_k}\right) = \quad (12)$$

$$= \left(-\frac{k}{\alpha^2}\frac{\partial^2 \rho_0}{\partial x_k^2} - \frac{k}{\alpha^2}\frac{\partial^2 \rho_1}{\partial x_k^2} - \frac{k}{\alpha^2}\frac{\partial^2 \rho_2}{\partial x_k^2} - \frac{k}{\alpha^2}\frac{\partial^2 \rho_3}{\partial x_k^2} + \alpha e^{-\alpha t}\frac{\partial \phi_k}{\partial x_k}\right) = -\frac{\partial \rho_0}{\partial t} - \frac{\partial \rho_1}{\partial t} - \frac{\partial \rho_2}{\partial t} - \frac{\partial \rho_3}{\partial t} = -\frac{\partial \rho}{\partial t}.$$

and we get the continuity equation

$$\frac{\partial \rho}{\partial t} + \frac{\partial(\rho u_k)}{\partial x_k} = 0. \quad (13)$$

We can eliminate velocities and write the continuity equation in another form

$$\frac{\partial \rho}{\partial t} - \frac{k}{\alpha^2}\frac{\partial^2 \rho}{\partial x_k \partial x_k} + \alpha e^{-\alpha t}\frac{\partial \phi_k}{\partial x_k} = 0. \quad (14)$$

## 4. Expressions for $\rho$ derivatives

We can express derivatives of $\rho_k$ in terms of velocities $u_k$. For example, expression (P1-15) reads now

$$\frac{\partial \rho_0}{\partial x_j} = -\frac{\alpha^2}{k}\rho u_{0j}, \quad (15)$$

and similarly

$$\frac{\partial \rho_1}{\partial x_j} = -\frac{\alpha^2}{k}\left(\rho u_{1j} - \alpha e^{-\alpha t}\delta_{1k}\phi_1\right). \quad (16)$$

Sum (15) and three expressions (16) and get

$$\frac{\partial \rho}{\partial x_j} = -\frac{\alpha^2}{k}\left(\rho u_j - \alpha e^{-\alpha t}\phi_j\right). \quad (17)$$

in full agreement with (11).

This means, that the field $\left(\rho u_i - \alpha e^{-\alpha t}\phi_i\right)$ posses potential. This potential is equal to $\left(-\frac{k}{\alpha^2}\rho\right)$.

We can also express second derivatives of $\rho_k$ in terms of velocities $u_k$ and their derivatives. For example, differentiation of (15) gives

$$\frac{\partial^2 \rho_0}{\partial x_i \partial x_j} = -\frac{\alpha^2}{k}\frac{\partial \rho}{\partial x_i}u_{0j} - \frac{\alpha^2}{k}\rho\frac{\partial u_{0j}}{\partial x_i}. \tag{18}$$

Pick expression for $\frac{\partial \rho}{\partial x_j}$ from (17) and insert it to (18)

$$\frac{\partial^2 \rho_0}{\partial x_i \partial x_j} = \left(\frac{\alpha^2}{k}\right)^2\left(\rho u_i - \alpha e^{-\alpha t}\phi_i\right)u_{0j} - \frac{\alpha^2}{k}\rho\frac{\partial u_{0j}}{\partial x_i}. \tag{19}$$

Similarly differentiation of (16) gives

$$\frac{\partial \rho_1}{\partial x_i \partial x_j} = -\frac{\alpha^2}{k}\frac{\partial \rho}{\partial x_i}u_{1j} - \frac{\alpha^2}{k}\left(\rho\frac{\partial u_{1j}}{\partial x_i} - \alpha e^{-\alpha t}\delta_{1k}\frac{\partial \phi_1}{\partial x_i}\right) = \tag{20}$$

$$= \left(\frac{\alpha^2}{k}\right)^2\left(\rho u_i - \alpha e^{-\alpha t}\phi_i\right)u_{1j} - \frac{\alpha^2}{k}\left(\rho\frac{\partial u_{1j}}{\partial x_i} - \alpha e^{-\alpha t}\delta_{1k}\frac{\partial \phi_1}{\partial x_i}\right).$$

Sum (19) and three expressions (20) and get

$$\frac{\partial \rho}{\partial x_i \partial x_j} = \left(\frac{\alpha^2}{k}\right)^2\left(\rho u_i - \alpha e^{-\alpha t}\phi_i\right)u_j - \frac{\alpha^2}{k}\left(\rho\frac{\partial u_j}{\partial x_i} - \alpha e^{-\alpha t}\frac{\partial \phi_j}{\partial x_i}\right). \tag{21}$$

Alternate (21) on the indices $i$, $j$

$$\left(\rho\frac{\partial u_j}{\partial x_i} - \alpha e^{-\alpha t}\frac{\partial \phi_j}{\partial x_i}\right) - \left(\rho\frac{\partial u_i}{\partial x_j} - \alpha e^{-\alpha t}\frac{\partial \phi_i}{\partial x_j}\right) + \frac{\alpha^2}{k}\alpha e^{-\alpha t}(\phi_i u_j - \phi_j u_i) = 0. \tag{22}$$

We shall use this identity (see (41)).

$$\frac{\partial \rho}{\partial x_i \partial x_j} = \left(\frac{\alpha^2}{k}\right)^2\left(\rho u_i u_j - \frac{\alpha}{2}e^{-\alpha t}\phi_i u_j - \frac{\alpha}{2}e^{-\alpha t}\phi_j u_i\right) - \frac{1}{2}\frac{\alpha^2}{k}\left[\rho\left(\frac{\partial u_j}{\partial x_i} + \frac{\partial u_i}{\partial x_j}\right) - \alpha e^{-\alpha t}\left(\frac{\partial \phi_j}{\partial x_i} + \frac{\partial \phi_i}{\partial x_j}\right)\right]. \tag{23}$$

Another interesting identity follows from (23) and (14)

$$\frac{k}{\alpha^2}\sum_{k=1}^{3}\left(\frac{\partial \rho}{\partial x_k \partial x_k} - \alpha e^{-\alpha t}\frac{\partial \phi_k}{\partial x_k}\right) = \frac{\partial \rho}{\partial t} = \sum_{k=1}^{3}\left(\frac{\alpha^2}{k}\left(\rho u_k - \alpha e^{-\alpha t}\phi_k\right)u_k - \rho\frac{\partial u_k}{\partial x_k}\right). \tag{24}$$

## 5. Current of momentum tensor and stresses

The expression for current of momentum tensor is (see [2] and [3]):

$$J_{1kl} = \int\limits_{-\infty}^{+\infty}\int\limits_{-\infty}^{+\infty}\int\limits_{-\infty}^{+\infty} \exp\left[-t\left(\alpha + \frac{k}{\alpha^2}\sum_{j=1}^{3}\omega_j^2\right)\right]\times \quad (25)$$

$$\times A_{\omega_1\omega_2\omega_3 100}\left(\frac{k}{2\pi\alpha}\right)^{3/2}\left(\left[-\frac{k}{\alpha}\frac{\omega_k}{\alpha} - \frac{\alpha\delta_{1k}}{\omega_1}\right]\left[-\frac{k}{\alpha}\frac{\omega_l}{\alpha} - \frac{\alpha\delta_{1l}}{\omega_1}\right] - \delta_{kl}\left[\frac{k}{\alpha} + \frac{\alpha^2\delta_{1l}}{\omega_1^2}\right]\right)\times$$

$$\times\left(\frac{2k}{\alpha}\right)^{1/2}\left(\frac{i\omega_1}{\alpha}\right)\prod_{j=1}^{j=3}\exp(i\omega_j x_j)\exp\left[-\frac{k}{2\alpha}\left(\frac{\omega_j}{\alpha}\right)^2\right]d\omega_1 d\omega_2 d\omega_3.$$

We see after, that simplification

$$\left(-\frac{k}{\alpha}\frac{\omega_k}{\alpha} - \frac{\alpha\delta_{1k}}{\omega_1}\right)\left(-\frac{k}{\alpha}\frac{\omega_l}{\alpha} - \frac{\alpha\delta_{1l}}{\omega_1}\right) - \delta_{kl}\left(\frac{k}{\alpha} + \frac{\alpha^2\delta_{1l}}{\omega_1^2}\right) = \quad (26)$$

$$= \left(\frac{k}{\alpha^2}\right)^2\omega_k\omega_l + \frac{\delta_{1k}}{\omega_1}\frac{k}{\alpha}\omega_l + \frac{\delta_{1l}}{\omega_1}\frac{k}{\alpha}\omega_k - \delta_{kl}\frac{k}{\alpha}.$$

cancel all negative degree terms.

$$J_{1kl} = e^{-\alpha t}\left[\left(\frac{k}{\alpha^2}\right)^2\frac{\partial^3\phi_1}{\partial x_1\partial x_k\partial x_l} - \delta_{1k}\frac{k}{\alpha}\frac{\partial\phi_1}{\partial x_l} - \delta_{1l}\frac{k}{\alpha}\frac{\partial\phi_1}{\partial x_k} + \delta_{kl}\frac{k}{\alpha}\frac{\partial\phi_1}{\partial x_1}\right]. \quad (27)$$

Recall (P1-14)

$$J_{0ij} = \left(\frac{k}{\alpha^2}\right)^2\frac{\partial^2\rho_0}{\partial x_i\partial x_j} + \frac{k}{\alpha}\rho_0\delta_{ij}. \quad (28)$$

Sum (28) and tree expressions (27)

$$J_{kl} = J_{0kl} + J_{1kl} + J_{2kl} + J_{3kl} = \left(\frac{k}{\alpha^2}\right)^2\frac{\partial^2\rho}{\partial x_k\partial x_l} - \frac{k}{\alpha}e^{-\alpha t}\left(\frac{\partial\phi_k}{\partial x_l} + \frac{\partial\phi_l}{\partial x_k}\right) + \delta_{kl}\frac{k}{\alpha}\rho = \quad (29)$$

$$= \left(\rho u_k - \alpha e^{-\alpha t}\phi_k\right)u_l - \left(\frac{k}{\alpha^2}\right)\left(\rho\frac{\partial u_l}{\partial x_k} - \alpha e^{-\alpha t}\frac{\partial\phi_l}{\partial x_k}\right) - \frac{k}{\alpha}e^{-\alpha t}\left(\frac{\partial\phi_k}{\partial x_l} + \frac{\partial\phi_l}{\partial x_k}\right) + \delta_{kl}\frac{k}{\alpha}\rho =$$

$$= \left[\rho u_k u_l - \frac{\alpha}{2}e^{-\alpha t}(\phi_k u_l + \phi_k u_l)\right] - \frac{1}{2}\frac{k}{\alpha^2}\left[\rho\left(\frac{\partial u_k}{\partial x_l} + \frac{\partial u_l}{\partial x_k}\right) - \alpha e^{-\alpha t}\left(\frac{\partial\phi_k}{\partial x_l} + \frac{\partial\phi_l}{\partial x_k}\right)\right] - \frac{k}{\alpha}e^{-\alpha t}\left(\frac{\partial\phi_k}{\partial x_l} + \frac{\partial\phi_l}{\partial x_k}\right) + \delta_{kl}\frac{k}{\alpha}\rho =$$

$$= \left[\rho u_k u_l - \frac{\alpha}{2}e^{-\alpha t}(\phi_k u_l + \phi_k u_l)\right] - \frac{1}{2}\frac{k}{\alpha^2}\left[\rho\left(\frac{\partial u_k}{\partial x_l} + \frac{\partial u_l}{\partial x_k}\right) + \alpha e^{-\alpha t}\left(\frac{\partial\phi_k}{\partial x_l} + \frac{\partial\phi_l}{\partial x_k}\right)\right] + \delta_{kl}\frac{k}{\alpha}\rho.$$

Tensor of stresses components are equal to

$$\sigma_{kl} = \rho u_k u_l - J_{kl} = \alpha e^{-\alpha t}\phi_k u_l + \frac{k}{\alpha^2}\left(\rho\frac{\partial u_l}{\partial x_k} + \alpha e^{-\alpha t}\frac{\partial \phi_k}{\partial x_l}\right) - \delta_{kl}\frac{k}{\alpha}\rho = \quad (30)$$

$$= \frac{\alpha}{2}e^{-\alpha t}(\phi_k u_l + \phi_l u_k) + \frac{1}{2}\frac{k}{\alpha^2}\left[\rho\left(\frac{\partial u_k}{\partial x_l} + \frac{\partial u_l}{\partial x_k}\right) + \alpha e^{-\alpha t}\left(\frac{\partial \phi_k}{\partial x_l} + \frac{\partial \phi_l}{\partial x_k}\right)\right] - \delta_{kl}\frac{k}{\alpha}\rho.$$

Recall, that in [1] we get the state equation of compressible viscous fluid with kinematic viscosity equal to $\nu = \frac{k}{\alpha^2}$. We see, that additional term is present in (30) - the first term. This term is proportional to $e^{-\alpha t}$ and therefore it is relatively small for big $t$.

## 6. Equations of movement

Now we consider another conservation law - equation of movement. The equation is:

$$\frac{\partial(\rho u_i)}{\partial t} + \frac{\partial(\rho u_i u_j)}{\partial x_j} - \frac{\partial \sigma_{ij}}{\partial x_j} + \alpha\rho u_i = 0. \quad (31)$$

We make following substitutions in (31) (all expressions are known from the previous text) :

$$\rho u_k = -\frac{k}{\alpha^2}\frac{\partial \rho}{\partial x_k} + \alpha e^{-\alpha t}\phi_k. \quad (32)$$

$$\rho u_k u_l - \sigma_{kl} = J_{kl}. \quad (33)$$

The result is:

$$\frac{\partial}{\partial t}\left(-\frac{k}{\alpha^2}\frac{\partial \rho}{\partial x_i} + \alpha e^{-\alpha t}\phi_i\right) + \quad (34)$$

$$+\frac{\partial}{\partial x_j}\left[\left(\rho u_i u_j - \frac{\alpha}{2}e^{-\alpha t}(\phi_i u_j + \phi_j u_i)\right) - \frac{1}{2}\frac{k}{\alpha^2}\left[\rho\left(\frac{\partial u_i}{\partial x_j} + \frac{\partial u_j}{\partial x_i}\right) + \alpha e^{-\alpha t}\left(\frac{\partial \phi_i}{\partial x_j} + \frac{\partial \phi_j}{\partial x_i}\right)\right] + \delta_{ij}\frac{k}{\alpha}\rho\right] +$$

$$+\alpha\rho u_i = 0.$$

We make further substitutions:

$$\frac{\partial \rho}{\partial t} = -\sum_{k=1}^{3}\frac{\partial(\rho u_k)}{\partial x_k}; \quad (35)$$

$$\frac{\partial \phi_i}{\partial t} = \frac{k}{\alpha^2}\left(\frac{\partial^2 \phi_i}{\partial x^2} + \frac{\partial^2 \phi_i}{\partial y^2} + \frac{\partial^2 \phi_i}{\partial z^2}\right); \quad (36)$$

$$\frac{\partial}{\partial x_j}\left(\delta_{ij}\frac{k}{\alpha}\rho\right) = \frac{k}{\alpha}\frac{\partial \rho}{\partial x_i} = -\alpha\left(\rho u_i - \alpha e^{-\alpha t}\phi_i\right) \qquad (37)$$

and get

$$\frac{k}{\alpha^2}\frac{\partial^2(\rho u_j)}{\partial x_i \partial x_j} - \alpha^2 e^{-\alpha t}\phi_i + \alpha e^{-\alpha t}\frac{\partial^2 \phi_i}{\partial x_j x_j} + \qquad (38)$$

$$+ \frac{\partial}{\partial x_j}\left\{\left(\rho u_i u_j - \frac{\alpha}{2}e^{-\alpha t}(\phi_i u_j + \phi_j u_i)\right) - \frac{1}{2}\frac{k}{\alpha^2}\left[\rho\left(\frac{\partial u_i}{\partial x_j} + \frac{\partial u_j}{\partial x_i}\right) + \alpha e^{-\alpha t}\left(\frac{\partial \phi_i}{\partial x_j} + \frac{\partial \phi_j}{\partial x_i}\right)\right]\right\} - \alpha\left(\rho u_i - \alpha e^{-\alpha t}\phi_i\right) +$$

$$+ \alpha \rho u_i = 0.$$

We can place all the expression under the brackets:

$$\frac{\partial}{\partial x_j}\left\{\frac{k}{\alpha^2}\frac{\partial(\rho u_j)}{\partial x_i} + \alpha e^{-\alpha t}\frac{\partial \phi_i}{\partial x_j} + \left(\rho u_i u_j - \frac{\alpha}{2}e^{-\alpha t}(\phi_i u_j + \phi_j u_i)\right) - \right. \qquad (39)$$

$$\left. - \frac{1}{2}\frac{k}{\alpha^2}\left[\rho\left(\frac{\partial u_i}{\partial x_j} + \frac{\partial u_j}{\partial x_i}\right) + \alpha e^{-\alpha t}\left(\frac{\partial \phi_i}{\partial x_j} + \frac{\partial \phi_j}{\partial x_i}\right)\right]\right\} = 0$$

and get

$$\frac{\partial}{\partial x_j}\left\{-\left(\rho u_i - \alpha e^{-\alpha t}\phi_i\right)u_j + \frac{k}{\alpha^2}\rho\frac{\partial u_j}{\partial x_i} + \alpha e^{-\alpha t}\frac{\partial \phi_i}{\partial x_j} + \left(\rho u_i u_j - \frac{\alpha}{2}e^{-\alpha t}(\phi_i u_j + \phi_j u_i)\right) - \right. \qquad (40)$$

$$\left. - \frac{1}{2}\frac{k}{\alpha^2}\left[\rho\left(\frac{\partial u_i}{\partial x_j} + \frac{\partial u_j}{\partial x_i}\right) + \alpha e^{-\alpha t}\left(\frac{\partial \phi_i}{\partial x_j} + \frac{\partial \phi_j}{\partial x_i}\right)\right]\right\} = 0.$$

We perform last simplifications and bring expression to the following form

$$\frac{\partial}{\partial x_j}\left\{\left(\frac{\alpha}{2}e^{-\alpha t}(\phi_i u_j - \phi_j u_i)\right) + \frac{1}{2}\frac{k}{\alpha^2}\left[\rho\left(\frac{\partial u_i}{\partial x_j} - \frac{\partial u_j}{\partial x_i}\right) + \alpha e^{-\alpha t}\left(\frac{\partial \phi_i}{\partial x_j} - \frac{\partial \phi_j}{\partial x_i}\right)\right]\right\} = 0. \qquad (41)$$

This equality is true because of identity (22).

In this way we checked, that variables $\rho$ and $v_i$ satisfy the set of conservation laws of classic hydrodynamics. They are not quite the set of equations of classic hydrodynamics of isothermal compressible fluid, because expression for stresses (30) contains additional term. This term is negligible in big times limit.

## 7. Burgers equation

We write equations of movement in slightly modified form (compare with [1]):

$$\frac{\partial u_i}{\partial t} + u_j \frac{\partial u_i}{\partial x_j} - \frac{1}{\rho}\frac{\partial \sigma_{ij}}{\partial x_j} + \alpha u_i = 0. \qquad (42)$$

We take expression for stresses in asymmetric form (see 30))

$$\sigma_{ij} = \alpha e^{-\alpha t}\phi_j u_i + \frac{k}{\alpha^2}\rho\frac{\partial u_i}{\partial x_j} + \frac{k}{\alpha}e^{-\alpha t}\frac{\partial \phi_j}{\partial x_i} - \delta_{ij}\frac{k}{\alpha}\rho. \qquad (43)$$

Perform substitution

$$\frac{\partial u_i}{\partial t} + u_j \frac{\partial u_i}{\partial x_j} - \frac{1}{\rho}\frac{\partial}{\partial x_j}\left(\alpha e^{-\alpha t}\phi_j u_i + \frac{k}{\alpha^2}\rho\frac{\partial u_i}{\partial x_j} + \frac{k}{\alpha}e^{-\alpha t}\frac{\partial \phi_j}{\partial x_i} - \delta_{ij}\frac{k}{\alpha}\rho\right) + \alpha u_i = 0 \qquad (44)$$

and get

$$\frac{\partial u_i}{\partial t} + u_j \frac{\partial u_i}{\partial x_j} - \frac{1}{\rho}e^{-\alpha t}\frac{\partial}{\partial x_j}\left(\alpha \phi_j u_i + \frac{k}{\alpha}\frac{\partial \phi_j}{\partial x_i}\right) - \qquad (45)$$

$$-\frac{1}{\rho}\frac{k}{\alpha^2}\frac{\partial \rho}{\partial x_j}\frac{\partial u_i}{\partial x_j} - \frac{k}{\alpha^2}\frac{\partial^2 u_i}{\partial x_j \partial x_j} + \frac{1}{\rho}\frac{k}{\alpha}\frac{\partial \rho}{\partial x_i} + \alpha u_i = 0.$$

We use expression for derivatives

$$\frac{\partial \rho}{\partial x_j} = -\frac{\alpha^2}{k}\left(\rho u_j - \alpha e^{-\alpha t}\phi_j\right) \qquad (46)$$

once again:

$$\frac{\partial u_i}{\partial t} + u_j \frac{\partial u_i}{\partial x_j} - \frac{1}{\rho}e^{-\alpha t}\frac{\partial}{\partial x_j}\left(\alpha \phi_j u_i + \frac{k}{\alpha}\frac{\partial \phi_j}{\partial x_i}\right) + \qquad (47)$$

$$+\frac{1}{\rho}\left(\rho u_j - \alpha e^{-\alpha t}\phi_j\right)\frac{\partial u_i}{\partial x_j} - \frac{k}{\alpha^2}\frac{\partial^2 u_i}{\partial x_j \partial x_j} - \frac{1}{\rho}\alpha\left(\rho u_i - \alpha e^{-\alpha t}\phi_i\right) + \alpha u_i = 0.$$

It follows

$$\frac{\partial u_i}{\partial t} + 2u_j \frac{\partial u_i}{\partial x_j} - \frac{k}{\alpha^2}\frac{\partial^2 u_i}{\partial x_j \partial x_j} - \frac{1}{\rho}e^{-\alpha t}\frac{\partial}{\partial x_j}\left(\alpha \phi_j u_i + \frac{k}{\alpha}\frac{\partial \phi_j}{\partial x_i}\right) - \qquad (48)$$

$$-\frac{1}{\rho}\alpha e^{-\alpha t}\phi_j\frac{\partial u_i}{\partial x_j} + \frac{1}{\rho}\alpha^2 e^{-\alpha t}\phi_i = 0.$$

Simple rearrangement of terms bring expression to its final form:

$$\frac{\partial u_i}{\partial t} + 2 u_j \frac{\partial u_i}{\partial x_j} - \frac{k}{\alpha^2} \frac{\partial^2 u_i}{\partial x_j \partial x_j} + \quad (49)$$

$$+ \frac{1}{\rho} e^{-\alpha t} \left[ \alpha^2 \phi_i - \frac{\partial}{\partial x_j} \left( \alpha \phi_j u_i + \frac{k}{\alpha} \frac{\partial \phi_j}{\partial x_i} \right) - \alpha \phi_j \frac{\partial u_i}{\partial x_j} \right] = 0.$$

As a result we get following equation for $u_i$ only. Similarly to [1], this equation strongly resembles Burgers equation. To get more usual form of Burgers equation we could perform substitution $t' = 2t$ and $v' = \frac{k}{2\alpha^2}$, but we omit this calculation.

We see from (49), that velocities $v_i$ satisfy the Burgers equation with some additional term. This term (the last term in (49)) is negligible in big times limit. Moreover, only this last term depend on $\rho$ variable. It is inversely proportional to $\rho$ and directly proportional to $\phi$ and its derivatives.

**DISCUSSION**

In this paper we proceed with investigation of connections between Fokker - Planck equation and continuum mechanics. We base upon expressions from our work [2], based upon the spectral decomposition of Fokker - Planck equation solution. In this decomposition we preserve only terms with the smallest degrees of damping - zero and first degree. We get expressions for density, velocities and stresses from potentials $\rho_0$, $\phi_i$. Potentials all satisfy the same diffusion equation. Starting from initial density and velocities, we could solve diffusion equations and find potentials. We find, that macroscopic parameters of Fokker-Planck flows, obtained in this way, satisfy the set of conservation laws of classic hydrodynamics. They are not quite the set of equations of classic hydrodynamics of isothermal compressible fluid, because expression for stresses (30) contains additional term. This term is negligible in big times limit. We proved also, that the velocities field alone, with potential, which satisfy diffusion equation, satisfy Burgers equation without mass forces - with some additional term. This term (the last term in (49)) is negligible in big times limit. Another approach to considered problem see in [5].

______________________________